\documentclass{article}
\usepackage{spconf,amsmath,graphicx}
\usepackage{amsfonts}
\usepackage[color=lime]{todonotes}
\usepackage{pgfplots}
\usepackage{pgfplotstable}
\usepackage{booktabs}
\usepackage{colortbl}
\usepackage{hyperref}
\usepgfplotslibrary{colorbrewer}

\newcommand{\x}{x}
\newcommand{\X}{X}
\newcommand{\M}{|X|}

\newcommand{\A}{\phi}
\newcommand{\Ah}{\widehat{\phi}}
\newcommand{\IF}{\psi}
\newcommand{\IFh}{\widehat{\psi}}
\newcommand{\GD}{\tau^g}
\newcommand{\GDh}{\bar{\tau}^g}
\newcommand{\z}{z}
\newcommand{\Dim}{d}
\newcommand{\Enc}{\operatorname{Enc}}
\newcommand{\Dec}{\operatorname{Dec}}
\renewcommand{\Re}{\operatorname{Re}}
\renewcommand{\Im}{\operatorname{Im}}
\newcommand{\GL}{\operatorname{GL}}

\newcommand{\birdsong}{\emph{Bird Audio Detection}}
\newcommand{\langid}{\emph{Spoken Language Identification}}
\newcommand{\librispeech}{\emph{LibriSpeech}}
\newcommand{\MUSAN}{\emph{MUSAN}}
\newcommand{\nsynthinstrument}{\emph{NSynthInstrument}}
\newcommand{\nsynthpitch}{\emph{NSynthPitch}}
\newcommand{\speechcommands}{\emph{Speech Commands}}
\newcommand{\TUT}{\emph{TUT Urban Acoustic Scenes 2018}}

\newcommand{\audiotovec}{\texttt{Audio2Vec}}
\newcommand{\autoencoder}{\texttt{AutoEncoder}}
\newcommand{\temporalgap}{\texttt{TemporalGap}}
\newcommand{\tripletloss}{\texttt{TripletLoss}}
\newcommand{\phaseprediction}{\texttt{PhasePredict}}

\newcommand{\projection}{\texttt{Projection}}
\newcommand{\untrained}{\texttt{Untrained}}

\newcommand{\multihead}{\texttt{MultiHead}}
\newcommand{\supervised}{\texttt{Supervised}}

\newcommand{\anglemean}[1]{\left\langle {#1} \right\rangle}

\pgfplotstableset{col sep=comma}
\pgfplotstableset{header=true}
\pgfplotstableset{every head row/.style={after row=\hline}}

\title{Learning audio representations via phase prediction}
\name{F\'elix de Chaumont Quitry, Marco Tagliasacchi, Dominik Roblek}
\address{Google Research \\ \small\texttt{\{fcq,mtagliasacchi,droblek\}@google.com}}

\begin{document}
\ninept
\maketitle
\begin{abstract}

We learn audio representations by solving a novel self-supervised learning task,
which consists of predicting the phase of the short-time Fourier transform from
its magnitude. A convolutional encoder is used to map the magnitude spectrum of
the input waveform to a lower dimensional embedding. A convolutional decoder is
then used to predict the instantaneous frequency (i.e., the temporal rate of
change of the phase) from such embedding. To evaluate the quality of the learned
representations, we evaluate how they transfer to a wide variety of downstream
audio tasks. Our experiments reveal that the phase prediction task leads to
representations that generalize across different tasks, partially bridging the
gap with fully-supervised models. In addition, we show that the predicted phase
can be used as initialization of the Griffin-Lim algorithm, thus reducing the
number of iterations needed to reconstruct the waveform in the time domain.

\end{abstract}

\section{Introduction}
\label{sec:intro}
The main goal of representation learning is to find a function that maps the
input data into a compact lower-dimensional embedding, which captures the
semantic properties of the input and can thus be re-used across different tasks.
In this respect, one conventional approach is to learn a fully-supervised
classifier that addresses simultaneously multiple classes~\cite{Hershey2017c},
and then use the activations computed by one of the final layers of the network
as embeddings. This approach suffers from two main shortcomings: first, it
relies on the availability of large labelled datasets (e.g.,
AudioSet~\cite{Gemmeke2017}); second, the learned representations might be
biased towards the specific classification task targeted during supervised
training. An alternative approach is to adopt unsupervised learning methods,
which have the potential to leverage the widespread availability of large
unlabelled datasets. Within this area, learning representations via
self-supervision has recently emerged as one viable approach, originally
explored in the field of language modeling~\cite{Mikolov2013a} and computer
vision~\cite{Zhang2016}, and more recently applied to
audio~\cite{Jansen2018,Tagliasacchi2019}. The key tenet of self-supervised
learning is the formulation of an auxiliary (or pretext) task, which does not
require access to labelled data. While solving the auxiliary task, the network
learns representations that can be potentially transferred to multiple
downstream tasks. The choice and the design of the task is particularly
important. The task should be neither too easy, so that the model is not able to
solve it exploiting low-level shortcuts, nor too hard, so that training can
converge.

In this paper we introduce a novel self-supervised task specifically tailored to
audio, which consists of predicting the phase of the short-time Fourier
transform (STFT) given the magnitude spectrum. This task is traditionally solved
as a by-product of the the well-known Griffin-Lim method~\cite{Griffin1984}, an
iterative algorithm that estimates the signal in the time domain from the STFT
magnitude. Instead, in this paper our primary interest is not the time-domain
reconstruction, but rather the audio representation that is learned by solving
the task. To some extent, our approach is reminiscent of the image colorization
task~\cite{Zhang2016} in computer vision, in which the three RGB channels are
reconstructed from a single grayscale channel. Similarly to~\cite{Zhang2016},
the task can be solved by leveraging the prior knowledge encapsulated in the
training data. An important aspect of our contribution is that we aim at
predicting the instantaneous frequency (IF), i.e., the temporal rate of change
of the instantaneous phase, rather than the raw phase. This is because the IF
tends to be smoother than the raw phase, and it can be therefore more easily
predicted given the inductive bias enforced by a convolutional decoder. Even so,
predicting the IF remains a challenging task, as it may still appear rather
random-like for the noisy parts of some sounds, such as fricative phonemes in
speech, or beats in music.

In our experiments we evaluate the quality of the learned representations by
training linear classifiers that receive the embeddings as input, targeting
multiple and heterogeneous downstream audio tasks, ranging from
music/speech/noise detection, to keywork spotting and speaker identification,
etc. We compare our results against other recently proposed self-supervised
learning methods, and we show that the representations are complementary to
those obtained with previous tasks.

\section{Related work}\label{sec:related_work}

Traditionally, audio representations are obtained by computing the output of
hand-crafted signal processing pipelines. These include MFCCs, PLP, LPC and
RASTA features, just to name a few examples. A more recent approach consists of
learning the representations directly from the data. For example,~\cite{Lee2009}
adopts an unsupervised learning approach based on deep belief networks to learn
representations applicable to speech and music. A wide variety of autoencoder
architectures have been explored, including denoising~\cite{Xu2017},
convolutional LSTM~\cite{Meyer2017} and sequence-to-sequence
autoencoders~\cite{Chung2016}. The model architecture that we adopt in this
paper is similar to an autoencoder, with the important difference that the
output (i.e., the instantaneous frequency) differs from the input (i.e., the
magnitude spectrum).

Self-supervised learning has been widely explored in the field of computer
vision, solving a wide variety of tasks ranging from predicting the relative
positions of image patches~\cite{Doersch2015}, to detecting image
rotations~\cite{Gidaris2018}, solving image inpainting~\cite{Pathak2016} or
image colorization~\cite{Zhang2016}. In contrast, fewer works applied
self-supervised learning to learn audio representations.
Speech2Vec~\cite{Chung2018b} and Audio2Vec~\cite{Tagliasacchi2019} learn,
respectively, speech and general audio representations, by reconstructing a
missing slice in the magnitude spectrogram. The temporal gap task
in~\cite{Tagliasacchi2019} estimates the temporal distance between two slices,
while the triplet loss is used in~\cite{Jansen2018}, creating anchor-positive
pairs by adding noise, shifting in time and/or frequency, and sampling temporal
slices. Our paper is mostly related to this field, and differs in the definition
of a novel self-supervised task. Note that representations learned by different
tasks can be complementary, and thus they can be potentially combined following
the approach in~\cite{Pascual2019}.

A problem closely related to phase prediction is that of spectrogram inversion,
where the goal is to obtain a phase which is consistent with a given input
magnitude spectrogram. This often comes as a second step in audio generative
models, such as~\cite{Vasquez2019} and~\cite{Donahue2019}. Spectrogram inversion
has traditionally been solved with the Griffin-Lim~\cite{Griffin1984} method,
but there are more recent alternatives such as~\cite{Shen2018}
and~\cite{Prenger2019}. Other audio generative models such as~\cite{Engel2019}
generate both the magnitude and the phase directly. Similarly to our
work,~\cite{Engel2019} also addresses the phase prediction problem by working in
the domain of the instantaneous frequency. Recently, a deep-learning based
method to refine the output of the Griffin-Lim algorithm has been proposed
in~\cite{Masuyama2019}.

\section{Method}\label{sec:method}

\begin{figure*}[t]
  \centering
  \includegraphics[trim=40 30 20 40,width=.75\textwidth]{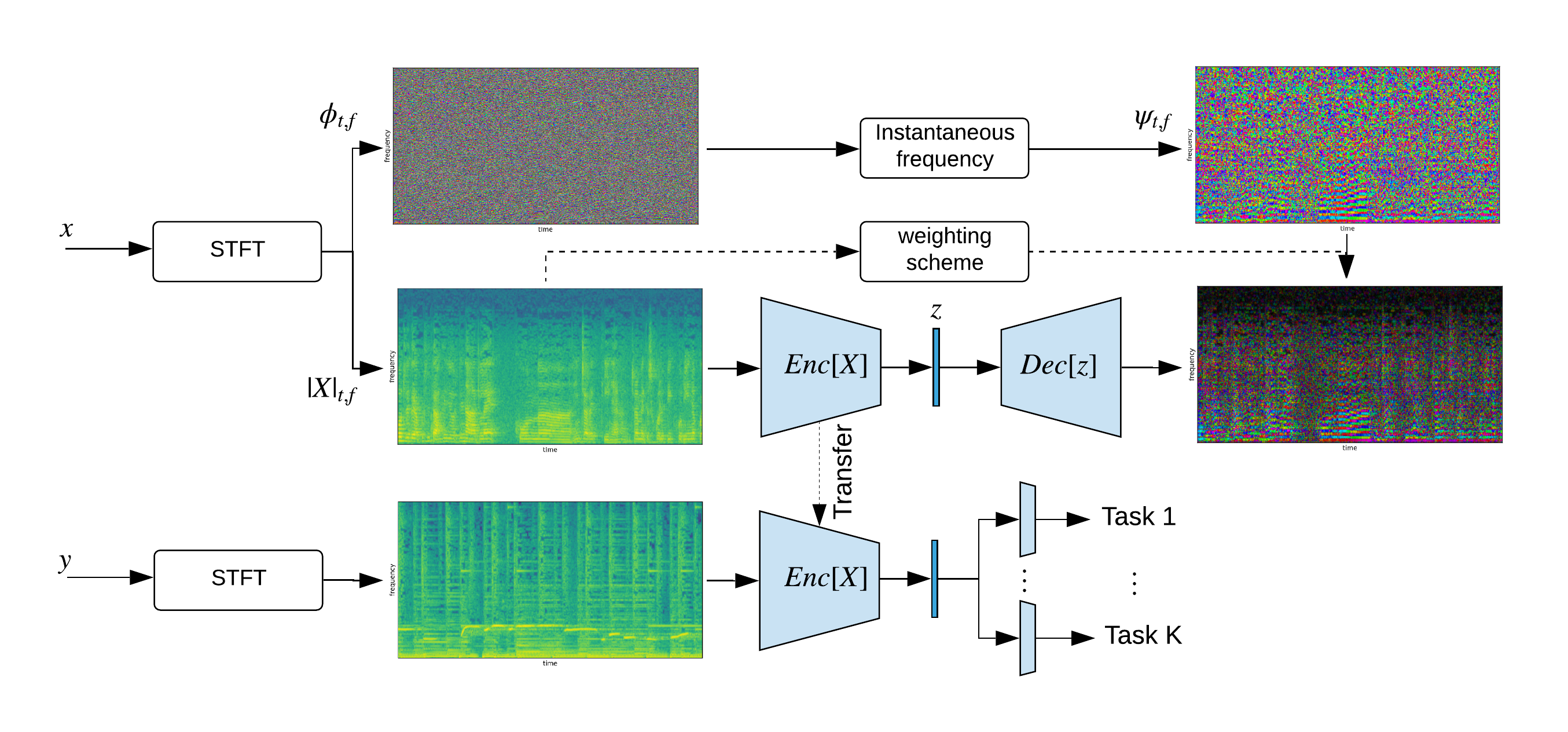}
  \caption{Model architecture of the proposed self-learning task.}
  \label{fig:model_architecture}
\end{figure*}

To learn useful audio representations with self-supervision, it is necessary to
devise a task that forces the model to extract some high-level features from the
input raw audio waveform. To this end, we propose a model that is designed much
like a spectrogram autoencoder, which receives as input the magnitude
spectrogram, but it has an important difference: instead of reconstructing the
magnitude itself, it aims at predicting the phase. An overview of the model is
illustrated in Figure~\ref{fig:model_architecture}. Note that after training the
self-supervised task, the features learned by the encoder can be transferred and
reused to address various downstream tasks by learning a simple head that maps
the embeddings to the target classes.

Formally, let $\x = \{x_1, x_2, \ldots, x_{n}\}$ denote an audio clip of $n$
samples in the time domain and $\X \in \mathbb{C}^{T \times F}$ the
corresponding complex spectrogram obtained computing the STFT, which consists of
$T$ temporal frames and $F$ frequency bins. Let $\z = \Enc(\M)$ denote a
$\Dim$-dimensional embedding computed by processing the input magnitude
spectrogram $\M$ with a convolutional encoder $\Enc()$, whose architecture is
detailed in Section~\ref{sec:experimental_results}. The phase prediction task is
then to predict the phase $\A = \arg(\X)$ using a decoder, i.e. $\Ah =
\Dec(\z)$. The phase is not corrected for the window hop delay, as is typically
the case when computing the STFT.

\textbf{Phase preprocessing}: We observed that attempting to directly
predicting the phase does not work. This is because the phase of an audio signal
appears to be noise-like, thus not matching the inductive bias of a
convolutional decoder, which favours the prediction of locally smooth signals.
For this reason, we first compute the instantaneous frequency (denoted $\IF$)
via phase differencing, as described in \cite{Boashash1992}, because it tends to
be significantly smoother along the temporal direction (compare for instance
$\A$ and $\IF$ in Figure~\ref{fig:model_architecture}). Another advantage of
this transformation is that it removes the global phase offset, which is
arbitrary and cannot be recovered from the magnitude alone. Phase differences
are then mapped to $[-\pi, \pi)$:

\begin{equation}
  \label{eq:instantaneous_frequency}
  \Delta\A_{t, f} = \A_{t+1, f} - \A_{t, f} \mod 2\pi
\end{equation}

We also experimented with using higher-order finite differences to better
approximate the continuous-time temporal derivative of the phase. This is
computed by first unwrapping the phase differences, smoothing the unwrapped
differences, and wrapping back. That is,

\begin{equation}
  \label{eq:smooth_inst_frequency}
  \begin{split}
    \Delta\A_{\cdot, f}^u &= \operatorname{unwrap}(\Delta\A_{\cdot, f}), \\
    \IF_{t, f} &= \frac{\Delta\A_{t-1, f}^u + \Delta\A_{t, f}^u}{2} \mod 2\pi.
  \end{split}
\end{equation}

This formulation has also the advantage of being a centered difference, so it is
naturally aligned with the corresponding amplitude along the temporal dimension.
Note that this is to be preferred to computing the centered difference directly
($(\A_{t+1, f} - \A_{t-1, f})/2$), since the latter only gives the correct value
modulo $\pi$.

\textbf{Phase loss}: The loss function captures how well the model predicts the
instantaneous frequency. Since the instantaneous frequency wraps around the unit
circle, we opted for the cosine loss instead of the mean square error:

\begin{equation}
  \label{eq:phase_loss}
  \mathcal{L}_{\IF} = \frac{1}{TF}\sum_{t, f} \left(1 - \cos(\IFh_{t, f} -
  \IF_{t, f})\right).
\end{equation}

\textbf{Phase error weighting}: In regions where the magnitude is close to zero,
the phase of the signal is not as meaningful. Therefore, attempting to
predict the instantaneous frequency in these cases does not efficiently exploit
the capacity of the model. To address this, we explored various alternatives for
weighting the error in the instantaneous frequency domain. We explored weighting
proportionally to $\M$ (as in~\cite{Afouras18}) and to $\sqrt{\M}$, and compared
this to using no weighting. In addition, having observed that the instantaneous
frequency is fairly smooth in the regions where it is well defined, we also
explored using the smoothness of $\IF$ itself to weight the errors made when
predicting $\IFh$. We computed smoothness as the inverse of the total variation
of the instantaneous phase, both in time and frequency:

\begin{equation}
  \label{eq:smoothness}
  S_{t, f} = \frac{1}{1 +
  \sum_{t',f'} |\IF_{t', f'} - \IF_{t, f} \mod 2\pi|}
\end{equation}

where $|t'-t| + |f'-f| = 1$.

\textbf{Hybrid autoencoder}: We also explored a variant in which we
simultaneously predict the instantaneous frequency and reconstruct the input
magnitude spectrogram. This results effectively in a self-supervised task which
is a hybrid between an magnitude autoencoder and our initial phase prediction
model. In this case we added to the loss function an additional term equal to
the mean square error between the input magnitude spectrogram and its
reconstruction.

\textbf{Output representation}: Given that $\IF$ wraps around the $[-\pi, \pi)$
interval, asking the model to produce $\IF$ directly requires it to sometimes
produce values which are discontinuous in time (i.e., ``jumping'' from $-\pi$ to
$\pi$, or vice-versa). For this reason, we also explored a second variant of an
hybrid autoencoder that aims at predicting simultaneously both the real and
imaginary parts ($\Re(\X)$ and $\Im(\X)$), which tend to be more continuous than
the phase.

\textbf{Audio reconstruction}: In our work we are primarily interested in
investigating phase prediction as a self-supervised task to learn new audio
representations. However, as a by-product, it is also possible to use the phase
estimated by the decoder together with the original magnitute spectrogram to
invert the STFT and reconstruct the signal in the time domain. In this way it is
possible to qualitatively appreciate the reconstruction by listening to the
audio samples. However, given that our model produces the instantaneous
frequency, we first need to integrate each frequency bin along time:

\begin{equation}
  \Ah_{t+1, f} = \Ah_{t, f} + \IFh_{t, f}.
\end{equation}

This means that we also need to estimate the initial phase offset of each
frequency bin, $\Ah_{0, f}$.

\textbf{Phase offset retrieval}: We define and compute group delay as the result
of phase differencing along the frequency dimension:

\begin{equation}
  \GD_{t, f} = \A_{t, f+1} - \A_{t, f} \mod 2\pi
\end{equation}

We also define the average of circular quantities as the angle of the barycenter
of the corresponding unit circle points:

\begin{equation}
  \anglemean{\vphantom{\sum}\theta_{t, f}}_t = \arg\left(\sum_t e^{i \theta_{t, f}}\right)
\end{equation}

We then observe that the average group delay, for fixed STFT parameters, is
narrowly centered around a single value $\GDh$. Therefore, it is sufficient to
choose the initial phase offset of each frequency bin so that the average group
delay of that frequency bin matches $\GDh$:

\begin{equation}
  \anglemean{\Ah_{t, f+1} - \Ah_{t, f}}_{t} = \GDh
\end{equation}

This is done by first choosing offsets equal to zero, computing the resulting
average group delays, and adjusting phase offsets:

\begin{equation}
  \Ah_{0, f+1} = \Ah_{0, f} +
  \GDh - \anglemean{\Ah_{t, f+1}^0 - \Ah_{t, f}^0}_{t},
\end{equation}
where we use $\Ah_{t, f}^0$, to denote the quantity intially obtained with
$\Ah_{0, f} = 0$. We still need to choose an arbitrary initial overall phase
value $\Ah_{0, 0}$, but this choice simply determines a global delay and it does
not affect the reconstruction in the time domain.

\section{Experimental results}

\label{sec:experimental_results}

\textbf{Audio front-end}: All our audio examples were sampled at 16~kHz, and
were converted to spectrograms via short-term Fourier transform (STFT), with a
window size of 25~ms, and a hop size of 10~ms. The DC bin was ignored. The
magnitude of the resulting spectrogram, when used either as input or as target
output, was normalized to zero mean and standard deviation equal to one. The
phase was converted to instantaneous frequency as described in
equation~\ref{eq:smooth_inst_frequency}.

\textbf{Audio tasks}: We evaluated the quality of the resulting embedding using
a variety of 8 downstream tasks, including speech- and music-related tasks as
well as auditory scene analysis tasks. More specifically, the {\speechcommands}
dataset (SPC)~\cite{Warden2018} contains 35 distinct spoken commands.
{\librispeech} (LSP)~\cite{Panayotov2015} contains 251 speakers reading
audiobooks, and is used here as a speaker identification task. The {\langid}
(LID) dataset~\cite{Tomasz2018} contains three languages (English, German, and
Spanish). The {\MUSAN} (MUS) dataset~\cite{Snyder2015} contains audio recordings
in three different environments: music, speech, and noise. We use the following
datasets from the DCASE2018 Challenge: {\birdsong}~\cite{Stowell2018} (BSD),
which is a binary detection task, and {\TUT}~\cite{Mesaros2018} (TUT), which
distinguishes between 10 different urban environments. Finally, we evaluate
performance on the NSynth dataset~\cite{Engel2017} using two tasks:
{\nsynthpitch} (NPI) attempts at estimating pitch out of 128 predefined pitch
levels, while {\nsynthinstrument} (NIF) contains recordings synthesized with 11
different instrument families. For all datasets, we respected the provided
train/test split, and report numbers obtained on the test set.

In all cases, the input audio waveforms were sliced into samples with a temporal
duration of 0.975s, resulting in spectrograms of size $T \times F$, where $T =
96$ and $F = 256$. Note that all selected tasks have a temporal granularity
which is compatible with the selected slice duration. In particular, we did not
include speech recognition tasks in our evaluation, since they typically require
a finer temporal granularity.

\textbf{Model architecture}: Our encoder $\Enc()$ uses a convolutional neural
network architecture with $L = 5$ layers, with respectively $[8, 16, 32, 64,
128]$ output channels. Each convolutional layer is followed by a max-pooling
layer dividing time and frequency dimensions by two, then ReLU activations, and
batch-normalization. The network ends with a global max-pooling layer and a
fully connected layer resulting in $\Dim = 128$-dimensional embeddings. The
decoder $\Dec()$ used in our experiments mirrors the encoder, though it should
be noted that using a more powerful decoder during training does not impact
inference when transferring the learned representations to new tasks. In
particular, we observed that when decoding into multiple outputs (e.g.,
magnitude and phase), it was mildly beneficial to use two separate decoders with
identical architectures, rather than a single decoder with two output channels.

\textbf{Baselines and benchmarks}: We compared our proposed method to the
following self-supervised methods previously proposed in the literature:
{\audiotovec} (in its two variants, CBoW and skip-gram) and
{\temporalgap}~\cite{Tagliasacchi2019}, {\tripletloss}~\cite{Jansen2018}, as
well as a regular {\autoencoder}. We also considered two additional baselines:
an {\untrained} encoder, based on the model architecture described above, but
with randomly initialized weights, and a $128$-dimensional random {\projection}
of the spectrogram. In all those cases, for each of the considered downstream
tasks, we train a single softmax layer that receives the embeddings as input.

Moreover, we include two fully-supervised benchmarks in our comparison: a
task-specific model, which consists of an encoder with the same architecture
described above followed by a single softmax layer, trained end-to-end
({\supervised}); a multi-head model, which consists of a shared encoder and
multiple task-specific softmax heads, trained jointly on all downstream tasks
({\multihead}).

\setlength{\tabcolsep}{5pt}
\begin{table}[t]
  \footnotesize
  \begin{filecontents*}{data/phaseweighting.csv}
weights,SPC,LID,LSP,MUS,TUT,BSD,NPI,NIF
{$\M$},0.2829906542056075,0.5333333333333333,0.7066809574848159,0.9584220165321982,0.5740740740740741,0.6908171861836563,0.6867013724562234,0.3140085186938003
{$\sqrt{\M}$},0.2641121495327103,0.4462962962962963,0.7699178277956413,0.9578280453398008,0.5679012345679012,0.6908171861836563,0.6901719514118946,0.3391702161224168
none,0.21813084112149533,0.5185185185185185,0.8160057163272597,0.9518883334158292,0.5950617283950618,0.6658242066835158,0.6859126045117526,0.3598359362675501
smooth,0.16953271028037384,0.4222222222222222,0.7588424437299035,0.9554521605702124,0.5555555555555556,0.686043246279135,0.6811011200504812,0.3744281432402587
\end{filecontents*}
\pgfplotstabletypeset[
      fixed, fixed zerofill,
      column type=c,
      columns/weights/.style={string type,column type=r},
    ]{data/phaseweighting.csv}
  \caption{Different weighting strategies for the phase loss}
  \label{tab:phase_weighting}
\end{table}

\setlength{\tabcolsep}{5pt}
\begin{table}[t]
  \footnotesize
  \begin{filecontents*}{data/hybrid.csv}
outputs,SPC,LID,LSP,MUS,TUT,BSD,NPI,NIF
{$\M$},0.4906542056074766,0.6055555555555555,0.7191854233654876,0.9505024006335692,0.5703703703703704,0.6711597865768043,0.4929010884997634,0.35849503076194983
{$\IF$},0.2829906542056075,0.5333333333333333,0.7066809574848159,0.9584220165321982,0.5740740740740741,0.6908171861836563,0.6867013724562234,0.3140085186938003
{$\M$} \& {$\IF$},0.4507476635514019,0.6037037037037037,0.7977849231868525,0.956788595753106,0.5802469135802469,0.7068239258635215,0.7027133617289794,0.3460324972393122
{$\Re$} \& {$\Im$},0.4247663551401869,0.562962962962963,0.7881386209360486,0.9439192199178341,0.5555555555555556,0.6731255265374895,0.6891465530840827,0.3359362675500868
\end{filecontents*}
\pgfplotstabletypeset[
      fixed, fixed zerofill,
      column type=c,
      columns/outputs/.style={string type,column type=r},
    ]{data/hybrid.csv}
  \caption{Hybrid autoencoders sometimes outperform both their counterparts}
  \label{tab:hybrid_autoencoders}
\end{table}

\setlength{\tabcolsep}{3pt}
\begin{table}[t]
  \centering
  \footnotesize
  \begin{filecontents*}{data/accuracy.csv}
model,SPC,LID,LSP,MUS,TUT,BSD,NPI,NIF
{\projection},0.2344859813084112,0.512962962962963,0.4262236513040372,0.8044844825025986,0.4765432098765432,0.6444818871103623,0.6661145291055371,0.2763842877425461
{\untrained},0.2424299065420561,0.43333333333333335,0.025366202215076808,0.873137652823838,0.4098765432098765,0.6467284470654311,0.35794289320082034,0.3023347531156334
{\autoencoder},0.4906542056074766,0.6055555555555555,0.7191854233654876,0.9505024006335692,0.5703703703703704,0.6711597865768043,0.4929010884997634,0.35849503076194983
{\phaseprediction},0.2829906542056075,0.5333333333333333,0.7066809574848159,0.9584220165321982,0.5740740740740741,0.6908171861836563,0.6867013724562234,0.3140085186938003
{\hybridautoencoder},0.4507476635514019,0.6037037037037037,0.7977849231868525,0.956788595753106,0.5802469135802469,0.7068239258635215,0.7027133617289794,0.3460324972393122
{\audiotovec} (cbow),0.5602803738317756,0.5370370370370371,0.8463737048946052,0.9764886403009454,0.6135802469135803,0.7169334456613311,0.592128095914182,0.4075563969080296
{\audiotovec} (skip),0.5423364485981308,0.4722222222222222,0.9071096820292962,0.9800524674553284,0.6382716049382716,0.7211457455770851,0.5635746963243414,0.40747752011358257
{\temporalgap},0.4441121495327103,0.4703703703703704,0.8999642729546267,0.9635697668663069,0.6271604938271605,0.7149677057006459,0.35936267550086765,0.38894147341852026
{\tripletloss},0.23485981308411216,0.4092592592592593,0.8903179707038228,0.9750037123199524,0.6333333333333333,0.7048581859028363,0.26194983435873165,0.38026502602934215
{\multihead},0.9010280373831776,0.6370370370370371,0.9964272954626652,0.9874276097609266,0.9074074074074074,0.7860151642796968,0.7863227638428775,0.5601041173686702
{\supervised},0.9325233644859812,0.6166666666666667,1.0,0.9939612928772954,0.9419753086419752,0.7843302443133952,0.8775043382236946,0.6009622968922543
\end{filecontents*}
\pgfplotstabletypeset[
      fixed, fixed zerofill,
      column type=c,
      columns/model/.style={string type,column type=r},
      every head row/.style={before row=\toprule, after row=\hline},
      every last row/.style={after row=\bottomrule},
      every row no 0/.style={before row={\rowcolor{red!10!white}}},
      every row no 1/.style={before row={\rowcolor{red!10!white}}},
      every row no 9/.style={before row={\rowcolor{cyan!10!white}}},
      every row no 10/.style={before row={\rowcolor{cyan!10!white}}},
    ]{data/accuracy.csv}
  \caption{Accuracy of linear classifiers trained on various embeddings}
  \label{tab:accuracy}
\end{table}

\begin{figure}[t]
  \centering
  \footnotesize
  \begin{tikzpicture}
    \begin{axis}[
        width=\linewidth,
        height=5cm,
        xlabel=Griffin-Lim iterations ($k$),
        ylabel=$\log{\left\Vert |\GL^k(\hat\X)| - \M\right\Vert_2}$,
        ymin=-1.6,
        ymax=0.1,
        legend pos=north east,
        legend cell align=left,
        legend style={nodes={scale=0.75, transform shape}},
        every axis plot/.append style={semithick},
        no markers,
        cycle list/Paired,
      ]
      \addplot table[x=steps, y=zero] {data/griffin_lim_numbers.csv};
      \addlegendentry{Zero phase}
      \addplot table[x=steps, y=random] {data/griffin_lim_numbers.csv};
      \addlegendentry{Random phase}
      \addplot table[x=steps, y=model_none] {data/griffin_lim_numbers.csv};
      \addlegendentry{{\phaseprediction} (no weights)}
      \addplot table[x=steps, y=model_smoothness] {data/griffin_lim_numbers.csv};
      \addlegendentry{{\phaseprediction} (smoothness)}
      \addplot table[x=steps, y=model_sqrtmag] {data/griffin_lim_numbers.csv};
      \addlegendentry{{\phaseprediction} ($\sqrt{\M}$ weights)}
      \addplot table[x=steps, y=model_mag] {data/griffin_lim_numbers.csv};
      \addlegendentry{{\phaseprediction} ($\M$ weights)}
    \end{axis}
  \end{tikzpicture}
  \caption{Log spectral convergence for various phase initializations.}
  \label{fig:griffin_lim_results}
\end{figure}
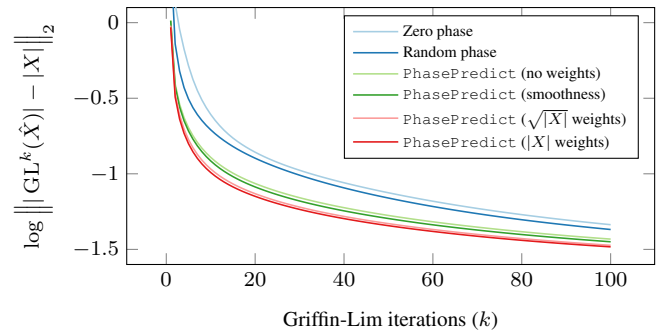

\textbf{Results}: First, we evaluate the impact of the weighting strategy
applied to phase prediction loss. Table~\ref{tab:phase_weighting} reports the
accuracy on all downstream tasks of the proposed self-supervised
{\phaseprediction} task. Overall, we observe that the results tend to be
relatively stable when varying the weighting strategy. For some tasks, e.g.,
{\speechcommands} and {\langid}, weighting based on the modulus of the
spectrogram leads to slightly higher accuracy. For other tasks, e.g.,
{\librispeech} and {\nsynthpitch}, better results are achieved when applying no
weighting at all. In the following, all results refer to the case of weighting
based on the modulus of the spectrogram.

Second, we evaluate how to combine the representations learned by the
{\phaseprediction} methods with those of a conventional autoencoder. To this
end, we explored two additional hybrid models: the first model has two separate
decoders producing, respectively, the magnitude and the phase; the second has
two separate decoders producing, respectively the real and imaginary parts of
the spectrogram. Table~\ref{tab:hybrid_autoencoders} summarizes these results,
showing that, for some downstream tasks the hybrid solution either improves
(e.g., for {\librispeech}, {\birdsong}, {\nsynthinstrument}) or matches
({\MUSAN}, {\TUT}, {\nsynthinstrument}), the accuracy attained when either
reconstructing the modulus or predicting the phase. Only for {\speechcommands}
is the accuracy slightly worse than that of a regular autoencoder. The last row
of Table~\ref{tab:hybrid_autoencoders} shows that predicting the real and
imaginary part leads to comparable, but slightly worse, performance on all
downstream tasks. Therefore, in the following we compare the hybrid autoencoder
that outputs the modulus and phase to other self-supervised methods.

Table~\ref{tab:accuracy} reports the accuracy attained when the embeddings
computed according to different self-supervised tasks are transferred to each of
the downstream tasks. As can be seen, among all the methods we have considered,
it is not possible to find a single self-supervised method that outperforms all
other methods on every task. Generally, the proposed {\phaseprediction} methods
achieves a level of accuracy comparable to other methods, performing
particularly well on {\nsynthpitch} and less so on {\speechcommands}. As already
investigated in~\cite{Pascual2019} in the context of learning representations
for speech-related tasks, different self-supervised tasks might learn different,
potentially complementary representations, and combining them generally leads to
richer representations that better generalize to downstream tasks. This is an
interesting research direction which is left to future work.

In addition to learning a representation that transfers to downstream tasks, the
estimated phase can be used in conjunction with the magnitude of the spectrogram
to recover the audio waveform in the time domain. As can be seen in
Figure~\ref{fig:griffin_lim_results}, the proposed method provides a good
initialization for the Griffin-Lim algorithm, thus reducing the number of
iterations needed to achieve a high-quality reconstruction. A qualitative
inspection of the reconstructed waveforms is also available
online\footnote{\url{https://google.github.io/phase-prediction/}}.

\section{Conclusions}\label{sec:conclusions}

In this paper we propose a novel audio-specific self-supervised task which is
able to learn representations in the absence of labelled data. The proposed task
consists of predicting the instantaneous frequency (i.e., the time-difference of
the phase) from the modulus of the STFT. Experimental results show that the
learned representations perform on a par with other recently proposed
self-supervised tasks, especially when combined with a traditional autoencoder
approach.

\bibliographystyle{IEEEbib}
\bibliography{references}

\end{document}